\documentclass[pra,showpacs,superscriptaddress,twocolumn]{revtex4}
\usepackage{graphicx,amsmath,amssymb,epsfig}

\begin{document}

\title{Non-Gaussian states produced by close-to-threshold optical parametric
oscillators: role of classical and quantum fluctuations}
\author{V. D'Auria}
\email{virginia.dauria@spectro.jussieu.fr}
\affiliation{Lab. Kastler Brossel, Universit\'{e} Pierre et Marie Curie, Ecole Normale Sup%
\'{e}rieure, CNRS, 4 place Jussieu, 75252 Paris, France}
\author{C. de Lisio}
\affiliation{CRS Coherentia CNR--INFM, I-80126 Napoli, Italia.}
\affiliation{Dip. di Scienze Fisiche, Univ. \textquotedblleft Federico
II\textquotedblright , Complesso Univ. Monte Sant'Angelo, I-80126 Napoli,
Italia.}
\author{A. Porzio}
\email{alberto.porzio@na.infn.it}
\affiliation{CNISM UdR Napoli Universit\`{a}, I-80126 Napoli, Italia.}
\author{S. Solimeno}
\affiliation{Dip. di Scienze Fisiche, Univ. \textquotedblleft Federico
II\textquotedblright , Complesso Univ. Monte Sant'Angelo, I-80126 Napoli,
Italia.}
\affiliation{CNISM UdR Napoli Universit\`{a}, I-80126 Napoli, Italia.}
\author{Javaid Anwar}
\affiliation{Phys. Dept., COMSATS (CIIT) Islamabad, Pakistan.}
\author{M. G. A.~Paris}
\email{matteo.paris@fisica.unimi.it}
\affiliation{Dip. di Fisica, Universit\`{a} degli studi di Milano, I-10133 Milano, Italia.}
\affiliation{CNISM UdR Milano Universit\`a, I-20133 Milano , Italia.}
\affiliation{ISI Foundation, I-10133 Torino, Italia.}

\begin{abstract}
Quantum states with non-Gaussian statistics generated by optical parametric
oscillators (OPO) with fluctuating parameters are studied by means of the
Kurtosis excess of the external field quadratures. The field generated is
viewed as the response of a nonlinear device to the fluctuations of laser
pump amplitude and frequency, crystal temperature and cavity detuning, in
addition to quantum noise sources. The Kurtosis excess has been evaluated
perturbatively up to the third order in the strength of the crystal
nonlinear coupling factor and the second order in the classical fluctuating
parameters. Applied to the device described in Opt. Expr. \textbf{13},
948-956 (2005) the model has given values of the Kurtosi excess in good
agreement with the measured ones.
\end{abstract}

\maketitle

\section{Introduction}

Non Gaussian (NG) resources (\textit{i.e. }NG states and/or operations) \cite%
{dj1,dj2,ngd1,ngd2,ngd3,ngd4} are required for realizing relevant quantum
information protocols, as for example entanglement distillation and swapping 
\cite{ds1,ds2,ds3,ds4}. It has been demonstrated that they may improve the
fidelity of teleportation \cite{Tom,IPS2a,IPS2b} and cloning \cite{nonGclon}
and NG states are more effective in revealing nonlocality \cite%
{nl1,nl2,nl3,nl4}. Thus the reliable generation of NG single-- and two--mode
states assumes a relevant role. Recently, few scheme to generate NG states
based on conditional de-Gaussification protocols have been proposed \cite%
{Tom,IPS2a,IPS2b} and realized \cite{IPSWenger,ipsb}. Moreover, it has been
proven that phase diffusing a squeezed vacuum state makes it a NG ones \cite%
{ngd2,ngd3,ngd4}.

Departures from Gaussian statistics have been observed (see Ref. \cite{OPEX}%
) in the outcomes for the field quadrature $X_{\theta }=\left( e^{-i\theta
}a+e^{i\theta }a^{\dagger }\right) /2$ outing a degenerate Optical
Parametric Oscillator (OPO). These deviations have been quantified by
measuring for different operating conditions the departure of the fourth
moment $\left\langle X_{\theta }^{4}\right\rangle $ from its Gaussian value.
In particular, it has been observed a dependence on the quadrature phase $%
\theta $ with the maximum departure always appearing in correspondence of
the anti-squeezed amplitude quadrature ($\theta =0$).

No doubt the observed deviations are due to a nonlinear response of the
device to Gaussian quantum and classical fluctuating parameters, such as
laser pump amplitude and phase, cavity length and OPO crystal temperature.
While in the linear analysis only the quantum input noise contributes to the
OPO\ output signal, by expanding the equation of motion to higher orders
classical fluctuations contribute as well to the output. As a result of the
multiplicative mixing of several noise sources, the signal looses its
Gaussian character. The weight of these sources in determining the deviation
from the Gaussian statistics was not clear at time of publication of Ref. 
\cite{OPEX}. The present paper is meant to present and discuss a consistent
theoretical framework to explain and interpret the observed experimental
evidences.

OPOs rely on parametric down--conversion: a strong \textit{pump} beam at
frequency $\omega _{p}$ interacts in a non--linear crystal with the vacuum
fields thus generating two beams, \textit{signal} and \textit{idler}, at
frequencies $\omega _{s}$ and $\omega _{i}$ respectively \cite%
{DK1,PK2,Reid-Drummond QCnondegenerate}. This mechanism is represented in
the Hamiltonian by the product of three field operators, $a_{p}$ (pump), $%
a_{s}$ (signal) $a_{i}$ (idler) and described dynamically by Langevin
equations. The model herein presented starts by including in the Graham and
Haken Langevin Equations (GHLE) different classical noise sources. Then,
expanding $a_{s,i}$ as power series in the strengths of the quantum and
classical fluctuations a hierarchy of Langevin equations is obtained in
which the field at a given order acts as source for the next one. This has
been done up to the third order thus culminating with Fig. \ref{exptoth}
showing an adequate agreement of the computed difference (\textit{Kurtosis
excess}) $K_{\theta }=\left( \left\langle X_{\theta }^{4}\right\rangle
-3\left\langle X_{\theta }^{2}\right\rangle ^{2}\right) /\left\langle
X_{\theta }^{2}\right\rangle ^{2}$ with the measured values $K_{\theta }$ of 
\cite{OPEX}.

A perturbative analysis of an ideal OPO was already developed in \cite%
{Chaturvedi Dechoum Drummond}: according to it nonlinear contributions
become comparable to the linear output only at a relative distance from the
threshold of about $\sim 10^{-6}$ while the data of Ref. \cite{OPEX} were
measured at a distance of $\ \sim 5\times 10^{-2}$ so signalling the
occurrence of more complex mechanisms. Hence, it has been essential to
account for contributions from different fluctuating parameters each
represented as a Gaussian process weighted by its standard deviations $g_{i}$%
. For the sake of generality the GHLE \cite{Graham} model has been developed
for a non degenerate OPO whereas the field moments have been calculated for
a degenerate one so to allow a direct comparison with \cite{OPEX}.

Different order fields have been represented as the convolution of the
previous order one with a $2\times 2$ matrix $\mathbf{G}$ which becomes
singular on approaching the threshold. This singularity is at the origin of
the enhancement of higher order effects in the critical region.

The paper is organized as follows. In Section \ref{s:LE} the extended GHLE
model for an OPO with fluctuating parameters is introduced and discussed.
The squeezing of the intracavity field is discussed in Sec. \ref{s:intra}.
Section \ref{s:extra} deals with the statistics of the quadratures outside
the cavity as measured by a finite bandwidth detector. The deviation from
the Gaussian statistics in terms of $K_{\theta }$ is analyzed by dwelling on
the agreement of the results provided by the model with the experimental
findings. Plots of $K_{\theta }$ vs. $\theta $ and $K_{\theta }$ for $\theta
=0$ vs. $E^{2}$ are reported. Section \ref{s:outro} closes the paper with
concluding remarks. Details about the linearization, together with few
analytical derivations are reported in the appendices.%
%%%%%%%%%%%%%%%%%%%%%%%

\section{Langevin equations\label{s:LE}}

Consider the set of the three OPO cavity modes $a_{k}$ ($a_{p}=e^{-i\omega
_{p}t-i\phi _{p}}a_{0}$ pump mode at frequency $\omega _{p}$ and phase $\phi
_{p}$, $a_{s}=e^{-i\omega _{s}t-i\frac{1}{2}\phi _{p}}a_{1}$ and $%
a_{i}=e^{-i\omega _{i}t-i\frac{1}{2}\phi _{p}}a_{2}$ respectively signal and
idler modes with $\omega _{p}=\omega _{s}+\omega _{i}$) whose mutual
interaction under the action of a driving field $e^{-i\omega _{p}t}\mathcal{E%
}$ is described by the Hamiltonian%
\begin{equation}
H_{int}=i\hbar \frac{\chi }{2}\left( a_{1}a_{2}a_{0}^{\dagger
}-a_{0}a_{1}^{\dagger }a_{2}^{\dagger }\right) +i\hbar \left( \mathcal{E}%
^{\ast }a_{0}-\mathcal{E}a_{0}^{\dagger }\right)
\label{Hamiltonian interaction}
\end{equation}%
where\textbf{\ }$\chi $ is\textbf{\ }the coupling parameter proportional to
the crystal second order susceptibility $\chi ^{\left( 2\right) }$. Since
real lasers are characterized by a field\textbf{\ }$\mathcal{E}=\epsilon
\left( 1+g_{\mu _{p}}\hat{\mu}_{p}\right) e^{-i\phi _{p}}$ of constant
amplitude $\epsilon $ modulated by a fluctuating factor $1+g_{\mu _{p}}\mu
_{p}\left( t\right) $ ($\left\langle \mu _{p}\right\rangle =0$) times a
phase factor $e^{-i\phi _{p}}$, where $\phi _{p}\left( t\right) $ is a
slowly diffusing phase, i.e. $\left\langle \left( \phi _{p}\left( t\right)
-\phi _{p}\left( t^{\prime }\right) \right) \right\rangle ^{2}=\Delta _{\ell
}\left\vert t-t^{\prime }\right\vert $ with $\Delta _{\ell }$ the laser
linewidth\cite{Reid-Drummond QCnondegenerate}.

The cavity modes are characterized by damping factors $\gamma _{k,M},\gamma
_{k,x}$ $\left( \gamma _{k}=\gamma _{k,M}+\gamma _{k,x}\right) $ due
respectively to the output mirror ($M$) and the other loss mechanisms (x:
crystal absorption and scattering, absorption of the two mirrors, etc.). The
evolution of the cavity mode operators can be described by the Graham-Haken
Langevin equations (GHLE) \cite{Graham}: 
\begin{align}
\dot{a}_{j}=& -\left( \gamma _{j}-i\nu _{j}-i\frac{1}{2}\dot{\phi}%
_{p}\right) a_{j}+\frac{\chi }{2}a_{0}a_{j^{\prime }}^{\dag }  \notag \\
& +e^{-ig_{\varpi _{p}}\phi _{p}/2}R_{j}\qquad \qquad j=1,2\text{ and }j\neq
j^{\prime }  \notag \\
\dot{a}_{0}=& -\left( \gamma _{0}-i\nu _{0}-i\dot{\phi}_{p}\right) a_{0}-%
\frac{\chi ^{\ast }}{2}a_{1}a_{2}+\epsilon \left( 1+g_{\mu _{p}}\mu
_{p}\right)  \notag \\
& +e^{ig_{\varpi _{p}}\phi _{p}}R_{0}  \label{Langevin initial}
\end{align}%
where $R_{k}\left( t\right) =\sqrt{2\gamma _{k,M}}b_{k,M}+\sqrt{2\gamma
_{k,x}}b_{k,x}$ takes into account the delta correlated vacuum fluctuations, 
$\left\langle b_{k,M,x}\left( t\right) b_{k,M,x}^{\dag }\left( t^{\prime
}\right) \right\rangle =\delta \left( t-t^{\prime }\right) $, entering the
OPO cavity. Modes are assumed to be slightly detuned by $\nu _{k}=\frac{\pi c%
}{L_{k}}\left[ L_{k}\frac{\omega _{k}}{\pi c}\right] -\omega _{k}$, with $%
L_{k}$ the OPO optical length at frequency $\omega _{k}$ and $\left[ x\right]
$ the closest integer to $x$.

In the following we will indicate by $\kappa _{k}=\gamma _{k}-i\nu
_{k}=\left\vert \kappa _{k}\right\vert e^{-i\psi _{k}}$ (with $\psi
_{1}=\psi _{2}$) the complex damping coefficients, by $\kappa =\left\vert
\kappa _{1}+\kappa _{2}\right\vert /2$ the mean decay rate and by $\tau
=\kappa t$ the time normalized to the cavity lifetime $\kappa ^{-1}$. A
caret will mark quantities normalized to $\kappa $ (e.g. $\hat{\epsilon}%
=\epsilon /\kappa $) and a tilde those such that the integral of their
correlation function (e.g. $\left\langle \hat{\mu}_{p}\left( \tau \right)
\mu _{p}\left( \tau ^{\prime }\right) \right\rangle =C_{\mu _{p}}\left( \tau
-\tau ^{\prime }\right) $) is equal to 1 (e.g. $\int_{-\infty }^{\infty
}C_{\mu _{p}}\left( \tau \right) d\tau =1$). In particular the Gaussian
delta-correlated process $d\phi _{p}/dt$\ will be replaced by $d\phi
_{p}/d\tau =g_{\varpi _{p}}\varpi _{p}$\ with $\left\langle \varpi
_{p}\left( \tau \right) \varpi _{p}\left( \tau ^{\prime }\right)
\right\rangle =\delta \left( \tau -\tau ^{\prime }\right) $\ and $g_{\varpi
_{P}}=\sqrt[4]{\left\langle \hat{\Delta}_{\ell }^{2}\right\rangle }$.

In real devices, beside the fluctuations related to classical noise of the
laser beam,\ the parameters $\nu _{k}$ and $\chi $ of Eqs. (\ref{Langevin
initial}) experience also the effects of mechanical vibrations. Residual
fluctuations of the cavity optical length $\delta L_{k},$ at frequency $%
\omega _{k},$ induce deviations $\delta \nu _{k}=-\left( \delta
L_{k}/L_{k}\right) \omega _{k}$ of the mode detunings from their average
values $\left\langle \nu _{k}\right\rangle $ $.$ Usually, an active control
guarantees that the standard deviation of $\delta \nu _{0}$ is a small
fraction of $\gamma _{0}$.

The parameter $\chi $ is proportional to the crystal susceptibility $\chi
^{\left( 2\right) }$ through the Boyd-Kogelnik function $H_{BK}\left( \sigma
,\varkappa ,\xi \right) $ \cite{Boyd} of the phase--matching factor $\sigma
, $ the focusing parameter $\xi $ and the crystal absorption $\varkappa $. $%
\sigma \left( T\right) $ depends on the crystal temperature $T$ through the
refractive indices at the interaction wavelengths.\ If the cavity
configuration is far from the concentric one the dependence of $\xi $ on the
cavity geometry fluctuations can be neglected. Under this assumption $\chi $
will be replaced in the system (\ref{Langevin initial}) by $\bar{\chi}%
e^{-i\phi _{\chi }}\left( 1+g_{T}\delta \hat{T}\right) $\textbf{\ }with $%
\bar{\chi}$ depending on the slow variations of T while $\phi _{\chi }$ is a
phase depending on the position of the beam waist with respect to the
crystal center. With an accurate alignment $\phi _{\chi }$ can be set equal
to 0. $g_{T}$ is defined by: 
\begin{equation*}
g_{T}=\sqrt{\left\langle \delta T^{2}\right\rangle }\frac{d\log H_{BK}}{dT}
\end{equation*}

In conclusion, the OPO analyzed in the following is characterized by four
classical fluctuating parameters $g_{\mu _{p}}\hat{\mu}_{p},g_{\varpi
_{p}}\varpi _{p},g_{\nu _{k}}\delta \hat{\nu}$ and $g_{T}\delta \hat{T}$,
where $\hat{\mu}_{p}$, $\varpi _{p}$, $\delta \hat{\nu}$, and $\delta \hat{T}
$ are Gaussian processes with unit standard deviations, and $g_{\mu
_{p}},g_{\varpi _{p}},g_{\nu _{k}}\left( =\frac{\left( \delta
L_{k}/L_{k}\right) }{\left( \delta L_{0}/L_{0}\right) }\frac{\omega _{k}}{%
\omega _{0}}g_{\nu _{0}}\right) ,g_{T}$ the corresponding weights. These
four terms, together with $g_{\chi }=|\bar{\chi}|/(2|\kappa _{0}\kappa
|)^{1/2}$, describing the non--linear interaction of strength $\bar{\chi}$ 
\cite{Chaturvedi Dechoum Drummond}, determine the OPO dynamics. For typical
operating conditions, ($g_{\chi }\simeq 10^{-6},$ $\kappa \simeq 10\div 20$
MHz, $\sqrt{\left\langle \Delta _{\ell }^{2}\right\rangle }\simeq 1\div 1000$
Hz, $\sqrt{\left\langle \delta T^{2}\right\rangle }\simeq 1\div 10$\textbf{\ 
}mK, and $\partial n/\partial T\approx 10^{-6}\div 10^{-4}$) $g_{\mu
_{p}},g_{\varpi _{p}},g_{\nu _{k}},g_{T}$ range in the respective intervals $%
\ 10^{-2}\div 10^{-1},10^{-4}\div 10^{-2},10^{-5}\div 10^{-1},10^{-5}\div
10^{-4}$.

The extended GHLE system (\ref{Langevin initial}) may now be written as 
\begin{align}
\dot{a}_{j}=& -\left( \hat{\kappa}_{j}-i\frac{g_{\varpi _{p}}\varpi _{p}}{2}%
+ig_{\nu _{j}}\delta \hat{\nu}\right) a_{j}  \notag \\
& +\sqrt{\frac{\left\vert \hat{\kappa}_{0}\right\vert }{2}}\left(
1+g_{T}\delta \hat{T}\right) g_{\chi }a_{0}a_{j^{\prime }}^{\dagger }  \notag
\\
& +e^{-ig_{\varpi _{p}}\phi _{p}/2}\hat{R}_{j}  \notag \\
\dot{a}_{0}=& -\left( \hat{\kappa}_{0}-ig_{\varpi _{p}}\varpi _{p}+ig_{\nu
_{0}}\delta \hat{\nu}\right) a_{0}  \notag \\
& -\sqrt{\frac{\left\vert \hat{\kappa}_{0}\right\vert }{2}}\left(
1+g_{T}\delta \hat{T}\right) g_{\chi }a_{1}a_{2}+\hat{\epsilon}\left(
1+g_{\mu _{p}}\hat{\mu}_{p}\right)  \notag \\
& +e^{ig_{\varpi _{p}}\phi _{p}}\hat{R}_{0}  \label{Langevin}
\end{align}%
a dot indicating derivatives with respect to $\tau $.

The OPO admits a threshold value for the amplitude $\epsilon =\epsilon
^{th}=\left\vert \kappa _{0}\right\vert \sqrt{\left\vert \kappa _{1}\kappa
_{2}\right\vert }/\left( 2\left\vert \bar{\chi}\right\vert \right) $. Below
threshold,\textbf{\ }the mode $a_{0}$ has a non--zero mean value $r_{0}$,
which is related to the driving field amplitude $\hat{\epsilon}$ ($\hat{%
\epsilon}=\hat{\kappa}_{0}r_{0}$). Therefore separating the average part $%
r_{0}$ from the fluctuating one $\delta a_{0}=r_{0}\alpha _{0}$ we put $%
a_{0}=r_{0}\left( 1+\alpha _{0}\right) $, where $\alpha _{0}=\rho
_{0}-i\varphi _{0}$. Conversely the modes $a_{j}$ have zero mean value and
will be expressed in terms of rescaled operators $a_{j}=r_{j}\alpha _{j}$,
with the $r_{j}$ defined in terms of $r_{0}$ 
\begin{equation}
\left\vert \hat{\kappa}_{j}\right\vert r_{j}^{2}=\left\vert \hat{\kappa}%
_{0}\right\vert r_{0}^{2}\;.  \label{btrj}
\end{equation}%
Passing now from the amplitudes $a_{k}$ to the scaled quantities $\alpha
_{k} $, the extended GHLE (\ref{Langevin}) is rewritten as:%
\begin{align}
\dot{\alpha}_{j}+\hat{\kappa}_{j}\alpha _{j}=& E\left\vert \hat{\kappa}%
_{j}\right\vert \alpha _{j^{\prime }}^{\dagger }+\,g_{\chi }\hat{N}_{\chi
_{j}}  \notag \\
& +\left\vert \hat{\kappa}_{j}\right\vert E\left( g_{\hat{T}}\delta \hat{T}%
+\alpha _{0}\right) \alpha _{j^{\prime }}^{\dagger }  \notag \\
& +i\left( g_{\varpi _{p}}\varpi _{p}/2-g_{\nu _{j}}\delta \hat{\nu}\right)
\alpha _{j}  \notag \\
& +\left\vert \hat{\kappa}_{j}\right\vert Eg_{\hat{T}}\delta \hat{T}\alpha
_{0}\alpha _{j^{\prime }}^{\dagger }  \notag \\
\dot{\alpha}_{0}+\hat{\kappa}_{0}\alpha _{0}=& -E\left\vert \hat{\kappa}%
_{0}\right\vert \alpha _{1}\alpha _{2}+\,g_{\chi }\hat{N}_{\chi _{0}}+\hat{%
\kappa}_{0}g_{\mu _{p}}\hat{\mu}_{p}  \notag \\
& +ig_{\varpi _{p}}\varpi _{p}-ig_{\nu _{0}}\delta \hat{\nu}+ig_{\varpi
_{p}}\varpi _{p}\alpha _{0}  \notag \\
& -ig_{\nu _{0}}\delta \hat{\nu}\alpha _{0}-E\left\vert \hat{\kappa}%
_{0}\right\vert g_{T}\delta \hat{T}\alpha _{1}\alpha _{2}\;,
\label{below threshold system}
\end{align}%
where 
\begin{equation}
E=\frac{g_{\chi }}{\sqrt{2\hat{\kappa}_{0}\hat{\kappa}_{1}\hat{\kappa}_{2}}}%
\,\hat{\epsilon}\,e^{i\psi _{0}}  \label{parameter E}
\end{equation}%
The above equations describe the dynamics of the fluctuating fields $\alpha
_{k}$ as responses to the classical noise sources $\hat{\mu}_{p},\varpi
_{p},\delta \hat{\nu},\delta \hat{T}$ and quantum terms 
\begin{align*}
\hat{N}_{\chi _{j}}& =\hat{R}_{j}~e^{-ig_{\varpi _{p}}\phi _{p}/2}/\left(
g_{\chi }r_{j}\right) \\
\hat{N}_{\chi _{0}}& =\hat{R}_{0}~e^{ig_{\varpi _{p}}\phi _{p}}/\left(
g_{\chi }r_{0}\right) ~.
\end{align*}%
The system (\ref{below threshold system}) may be solved perturbatively upon
expanding the field amplitudes as 
\begin{equation}
\alpha _{k}=\sum_{m=1}^{3}\alpha _{k}^{\left( m\right) }.  \label{expansion}
\end{equation}%
The first-order terms ($m=1$) correspond to the linearized system. $\alpha
_{k}^{\left( m\geq 2\right) }$\textbf{\ }are generated by non linear sources 
$s_{k}^{\left( m\right) }$\ made of a quantum contribution, proportional to $%
g_{\chi }^{m}$,\ and of mixed terms involving products of quantum and
classical fluctuations of the type $g_{\chi }^{m-1}g_{i}$\ and $g_{\chi
}^{m-2}g_{i}g_{j}$. Any\ $s_{k}^{\left( m\right) }$ involves fields
calculated up to the ($m-1$)--th order. By substituting the $\alpha _{k}$
expansion into (\ref{below threshold system}) and grouping the terms
corresponding to the same perturbative order,\ each $\alpha _{k}^{\left(
m\right) }$ can be calculated as the convolution of the components $%
G_{kk^{\prime }}$ of the Green's matrix (see Eq. (\ref{components matrix
Gopo bt})) relative to the linearized system, with $s_{k}^{\left( m\right) }$
and $s_{k^{\prime }}^{\left( m\right) \dagger }$.

For a degenerate (the signal and idler fields collapse into a single field
in this case) and tuned ($\bar{\nu}_{k}=0$) OPO the extended GHLE (\ref%
{below threshold system}) reduces to 
\begin{align}
\dot{\alpha}+\hat{\kappa}\alpha =& E\alpha ^{\dagger }+g_{\chi }\hat{N}%
_{\chi }+E\left( g_{\tilde{T}}\delta \tilde{T}+\alpha _{0}\right) \alpha
^{\dagger }  \notag \\
& +i\left( \frac{1}{2}g_{\varpi _{p}}\varpi _{p}-g_{\nu _{j}}\delta \hat{\nu}%
\right) \alpha +Eg_{\tilde{T}}\delta \hat{T}\alpha _{0}\alpha ^{\dagger } 
\notag \\
\ \dot{\alpha}_{0}+\hat{\kappa}_{0}\alpha _{0}=& \,g_{\chi }\hat{N}_{\chi
_{0}}+\hat{\kappa}_{0}g_{\mu _{p}}\hat{\mu}_{p}+ig_{\varpi _{p}}\varpi
_{p}-ig_{\nu _{0}}\delta \hat{\nu}  \notag \\
& -\frac{E}{2}\left\vert \hat{\kappa}_{0}\right\vert \alpha ^{2}+ig_{\varpi
_{p}}\varpi _{p}\alpha _{0}-ig_{\nu _{0}}\delta \hat{\nu}\alpha _{0}  \notag
\\
& -E\left\vert \hat{\kappa}_{0}\right\vert g_{T}\delta \hat{T}\alpha ^{2}\;,
\label{below threshold degenerate}
\end{align}%
where the parameter%
\begin{equation*}
E=\frac{2g_{\chi }}{\sqrt{2\hat{\kappa}_{0}}}\hat{\epsilon}=e^{i\psi _{0}/2}%
\frac{\hat{\epsilon}}{\hat{\epsilon}^{th}}=e^{i\psi _{0}/2}\left\vert
E\right\vert
\end{equation*}%
now represents the excitation $\hat{\epsilon}$ normalized to the threshold $%
\hat{\epsilon}^{th}=\sqrt{\left\vert \hat{\kappa}_{0}\right\vert /2}/g_{\chi
}$ while $\hat{N}_{\chi }(\tau )$, $\hat{N}_{\chi _{0}}(\tau )$ and $\varpi
_{p}\left( \tau \right) $ are delta-correlated processes, 
\begin{eqnarray}
\left\langle \hat{N}_{\chi }\left( \tau \right) \hat{N}_{\chi }^{\dagger
}\left( \tau ^{\prime }\right) \right\rangle &=&\frac{4}{\left\vert
E\right\vert ^{2}}\delta \left( \tau -\tau ^{\prime }\right) ~,  \notag \\
\left\langle \hat{N}_{\chi _{0}}\left( \tau \right) \hat{N}_{\chi
_{0}}^{\dagger }\left( \tau ^{\prime }\right) \right\rangle &=&\frac{%
4\left\vert \hat{\kappa}_{0}\right\vert ^{2}}{\left\vert E\right\vert ^{2}}%
\delta \left( \tau -\tau ^{\prime }\right) ~,  \label{noise correlations}
\end{eqnarray}%
The correlation times for $\hat{\mu}_{p}$ are typically of the order of $%
0.2\div 1$ $\mu s$, while those for $\delta \hat{\nu}$ and $\delta \hat{T}$
are of the order of $ms$ and $s$ respectively. Although $\delta \hat{\nu}$
and $\delta \hat{T}$ can be treated adiabatically, we have preferred to
treat the noise sources in a unified fashion.

\subsection{Nonlinear terms for a degenerate OPO\label{s:nonlin}}

Expanding Eqs. (\ref{below threshold system}) up to the third order the
fields $\alpha ^{\left( m\right) }$ (signal=idler) and $\alpha _{0}^{\left(
m\right) }$ (pump) of (\ref{expansion}), represented in the vector form $%
\mathbf{\alpha }^{\left( m\right) }\mathbf{=}\left( \alpha ^{\left( m\right)
}\mathbf{,}\alpha ^{\left( m\right) \dag }\right) ^{T},\mathbf{\alpha }%
_{0}^{\left( m\right) }=\left( \alpha _{0}^{\left( m\right) }\mathbf{,}%
\alpha _{0}^{\left( m\right) \dag }\right) ^{T},$ are given by%
\begin{eqnarray*}
\mathbf{\alpha }^{\left( m\right) }\left( t\right) &=&\hspace{-1.35cm}%
\int^{t}\mathbf{G}\left( t-\tau \right) \mathbf{\cdot s}^{(m)}(\tau )d\tau \\
\mathbf{\alpha }_{0}^{\left( m\right) }\left( t\right) &=&\hspace{-1.35cm}%
\int^{t}\mathbf{G}_{0}\left( t-\tau \right) \mathbf{\cdot s}_{0}^{(m)}(\tau
)d\tau
\end{eqnarray*}%
with $\mathbf{G}$ and $\mathbf{G}_{0}$ defined by Eq. (\ref{Green time}),
while the signal $\mathbf{s}^{(m)}$ and pump $\mathbf{s}_{0}^{\left(
m\right) }$ sources read respectively:%
\begin{align}
\mathbf{s}^{(1)}(\tau )& =g_{\chi }\mathbf{N}_{\chi }(\tau )  \notag \\
\mathbf{s}^{(2)}(\tau )& =\mathbf{B}^{(1)}(\tau )\cdot \mathbf{\alpha }%
^{(1)}(\tau )  \notag \\
\mathbf{s}^{(3)}(\tau )& =B^{(2)}\mathbf{\alpha }^{(1)\dagger }(\tau ) 
\notag \\
& \hspace{-1.35cm}\int^{\tau }\left( g_{\chi }^{2}\delta B^{\left( 2\right)
}\left( \tau -\tau ^{\prime }\right) \mathbf{1}+\mathbf{B}^{\left(
1,1\right) }\left( \tau -\tau ^{\prime }\right) \right) \cdot \mathbf{\alpha 
}^{\left( 1\right) }\;\left( \tau ^{\prime }\right) d\tau ^{\prime }
\label{nonlinear source signal}
\end{align}%
and 
\begin{align}
\mathbf{s}_{0}^{(1)}(\tau )=& g_{\chi }\mathbf{\hat{N}}_{\chi _{0}}(\tau )+i%
\left[ g_{\varpi _{p}}\varpi _{p}(\tau )-g_{\nu _{0}}\delta \hat{\nu}(\tau )%
\right] \mathbf{1}_{-}  \notag \\
& +\hat{\kappa _{0}}g_{\mu _{p}}\hat{\mu}_{p}(\tau )  \notag \\
\mathbf{s}_{0}^{\left( 2\right) }\left( \tau \right) =& i\left[ g_{\varpi
_{p}}\varpi _{p}\alpha _{0}-ig_{\nu _{0}}\delta \hat{\nu}\alpha _{0}\right] 
\mathbf{1}_{-}-\frac{1}{2}E\hat{\kappa}_{0}\mathbf{\alpha }^{\left( 1\right)
2}  \label{nonlinear sources pump}
\end{align}%
with $\mathbf{1}_{-}=\left( 1,-1\right) ^{T}$ and $\mathbf{B}^{\left(
1\right) }$%
\begin{equation}
\mathbf{B}^{\left( 1\right) }=\left[ 
\begin{array}{ll}
i\left( \frac{1}{2}g_{\varpi _{p}}\varpi _{p}-g_{\nu }\delta \hat{\nu}\right)
& E\left( \alpha _{0}^{\left( 1\right) }+g_{T}\delta \hat{T}\right) \\ 
E^{\ast }\left( P\alpha _{0}^{\left( 1\right) \dagger }+g_{T}\delta \hat{T}%
\right) & -i\left( \frac{1}{2}g_{\varpi _{p}}\varpi _{p}-g_{\nu }\delta \hat{%
\nu}\right)%
\end{array}%
\right] ~,  \label{B1}
\end{equation}%
with $P$ the permutation operator. Note that $\delta \hat{T}$ and $\alpha
_{0}^{\left( 1\right) }$ appear in the off--diagonal terms while $\varpi
_{p}\ $and $\delta \hat{\nu}\ $are in the diagonal ones.

For a tuned OPO $\left( \psi =\psi _{0}=\phi _{\chi }=0\right) $ $B^{\left(
2\right) }$, $\delta B^{\left( 2\right) }$ and $\mathbf{B}^{\left(
1,1\right) }$, of Eq. (\ref{nonlinear source signal}--c), are given by 
\begin{align*}
B^{\left( 2\right) }=E\left\langle \alpha _{0}^{\left( 2\right)
}\right\rangle & =-\frac{g_{\chi }^{2}}{2\left( 1-E^{2}\right) }-g_{\varpi
_{p}}\frac{E}{\hat{\kappa}_{0}}-g_{\nu }^{2}\frac{E}{\hat{\kappa}_{0}^{2}} \\
\delta B^{\left( 2\right) }\left( \tau -\tau ^{\prime }\right) & =-E^{2}\hat{%
\kappa}_{0}G_{0}\left( \tau -\tau ^{\prime }\right) \sigma _{a\alpha
^{\dagger }}^{\left( 1,1\right) }\left( \tau -\tau ^{\prime }\right) \\
\mathbf{B}^{\left( 1,1\right) }\left( \tau -\tau ^{\prime }\right) &
=\left\langle \mathbf{B}^{\left( 1\right) }\left( \tau \right) \cdot \mathbf{%
G}\left( \tau -\tau ^{\prime }\right) \cdot \mathbf{\mathbf{B}}^{\left(
1\right) }\left( \tau ^{\prime }\right) \right\rangle .
\end{align*}%
with $\sigma _{a\alpha ^{\dagger }}^{\left( 1,1\right) }\left( \tau -\tau
^{\prime }\right) =\left\langle \alpha ^{\left( 1\right) }\left( \tau
\right) \alpha ^{\left( 1\right) \dagger }\left( \tau ^{\prime }\right)
\right\rangle $. To be consistent with the above hierarchy, all moments $%
\left\langle s^{\left( m\right) }s^{\left( m\right) }\right\rangle $ must
satisfy the inequalities $\left\langle s^{\left( 1\right) }s^{\left(
1\right) }\right\rangle >\left\langle s^{\left( 2\right) }s^{\left( 2\right)
}\right\rangle >\left\langle s^{\left( 3\right) }s^{\left( 3\right)
}\right\rangle $. Close to the critical point, $\left\langle s^{\left(
1\right) }s^{\left( 1\right) }\right\rangle $, $\left\langle s^{\left(
2\right) }s^{\left( 2\right) }\right\rangle $ and $\left\langle s^{\left(
3\right) }s^{\left( 3\right) }\right\rangle $ are of the order of $O\left(
g_{\chi }^{2}\right) $,$\ O\left( g_{\chi }^{2}g_{i}^{2}\left( 1-E\right)
^{-1}\right) $ and $O\left( g_{\chi }^{2}g_{i}^{4}\left( 1-E\right)
^{-2}\right) $ respectively, so that the above condition implies that the
approximation maintains its validity up to $1-E>g_{i}^{2}$.

Since the linear source for the down--converted beam (see Eq. (\ref%
{nonlinear source signal}--a)) contains uniquely the quantum noise term, the
effects of the classical fluctuations can be analyzed only going beyond the
linear approximation.

\section{Intracavity field\label{s:intra}}

The nonlinear contribution to the intracavity field is represented by the
averaged tensor product with respect to the different fluctuating parameters 
$g_{\iota }N_{\iota }$ ($g_{\chi }N_{\chi }$, $g_{\chi }N_{\chi _{0}}$, $%
g_{\mu _{p}}\hat{\mu}_{p}$, $g_{\varpi _{P}}\varpi _{P}$, $g_{\nu }\delta 
\hat{\nu}$, and $g_{T}\delta \hat{T}$), 
\begin{align*}
\boldsymbol{\sigma }^{\left( NL\right) }\left( \tau \right) & =\left\langle 
\mathbf{\alpha }\left( \tau \right) \mathbf{\alpha }^{T}\left( 0\right)
\right\rangle -\left\langle \mathbf{\alpha }^{\left( 1\right) }\left( \tau
\right) \mathbf{\alpha }^{\left( 1\right) T}\left( 0\right) \right\rangle  \\
& =\boldsymbol{\sigma }^{\left( 2,2\right) }\left( \tau \right) +\boldsymbol{%
\sigma }^{\left( 3,1\right) }\left( \tau \right) +\boldsymbol{\sigma }%
^{\left( 1,3\right) }\left( \tau \right)  \\
& =g_{\chi }^{2}\sum_{\iota }g_{\iota }^{2}\boldsymbol{\sigma }_{\iota
}^{\left( NL\right) }\left( \tau \right) 
\end{align*}%
Relevant $\boldsymbol{\sigma }^{\left( m,n\right) }\left( \tau \right) $ are
explicitly given by:%
\begin{widetext}%
\begin{eqnarray*}
\boldsymbol{\sigma }^{\left( 2,2\right) }\left( \tau \right)
&=&\int_{-\infty }^{\tau }d\tau ^{\prime }\int_{-\infty }^{0}d\tau ^{\prime
\prime }\mathbf{G}\left( \tau -\tau ^{\prime }\right) \cdot \left\langle 
\mathbf{B}^{\left( 1\right) }\left( \tau ^{\prime }\right) \cdot \boldsymbol{%
\sigma }^{\left( 1,1\right) }\left( \tau ^{\prime }-\tau ^{\prime \prime
}\right) \cdot \mathbf{B}^{\left( 1\right) T}\left( \tau ^{\prime \prime
}\right) \right\rangle \cdot \mathbf{G}\left( -\tau ^{\prime \prime }\right)
\\
\boldsymbol{\sigma }^{\left( 3,1\right) }\left( \tau \right)
&=&\int_{-\infty }^{0}d\tau \mathbf{G}\left( \tau -\tau ^{\prime }\right)
\cdot \left( B^{\left( 2\right) }\left[ 
\begin{array}{cc}
0 & 1 \\ 
1 & 0%
\end{array}%
\right] \right. \left. +\int_{-\infty }^{\tau }\left( \delta B^{\left(
2\right) }\left( \tau -\tau ^{\prime }\right) \mathbf{1}+\mathbf{B}^{\left(
1,1\right) }\left( \tau -\tau ^{\prime }\right) \right) \cdot \boldsymbol{%
\sigma }^{\left( 1,1\right) }\left( \tau ^{\prime }\right) d\tau ^{\prime
}\right)
\end{eqnarray*}%
\end{widetext}

Of particular interest is the variance $\left\langle X_{\pi /2}^{\left(
NL\right) 2}\right\rangle _{\iota }$ normalized to $\left\langle X_{\pi
/2}^{2}\right\rangle $, \textit{i.e.} the weight, normalized to $g_{\iota
}^{2}$, of the nonlinear correction to the squeezed variance:%
\begin{multline*}
\lambda _{\iota }^{NL}=\frac{\left\langle X_{\pi /2}^{\left( NL\right)
2}\right\rangle _{\iota }}{\left\langle X_{\pi /2}^{2}\right\rangle } \\
=\frac{-\boldsymbol{\sigma }_{\iota aa}^{\left( NL\right) }\left( 0\right) +%
\boldsymbol{\sigma }_{\iota aa^{\dagger }}^{\left( NL\right) }\left(
0\right) +\boldsymbol{\sigma }_{\iota a^{\dagger }a}^{\left( NL\right)
}\left( 0\right) -\boldsymbol{\sigma }_{\iota a^{\dagger }a^{\dagger
}}^{\left( NL\right) }\left( 0\right) }{4\left\langle X_{\pi
/2}^{2}\right\rangle }
\end{multline*}

\begin{figure}[h]
\includegraphics[width=0.48\columnwidth]{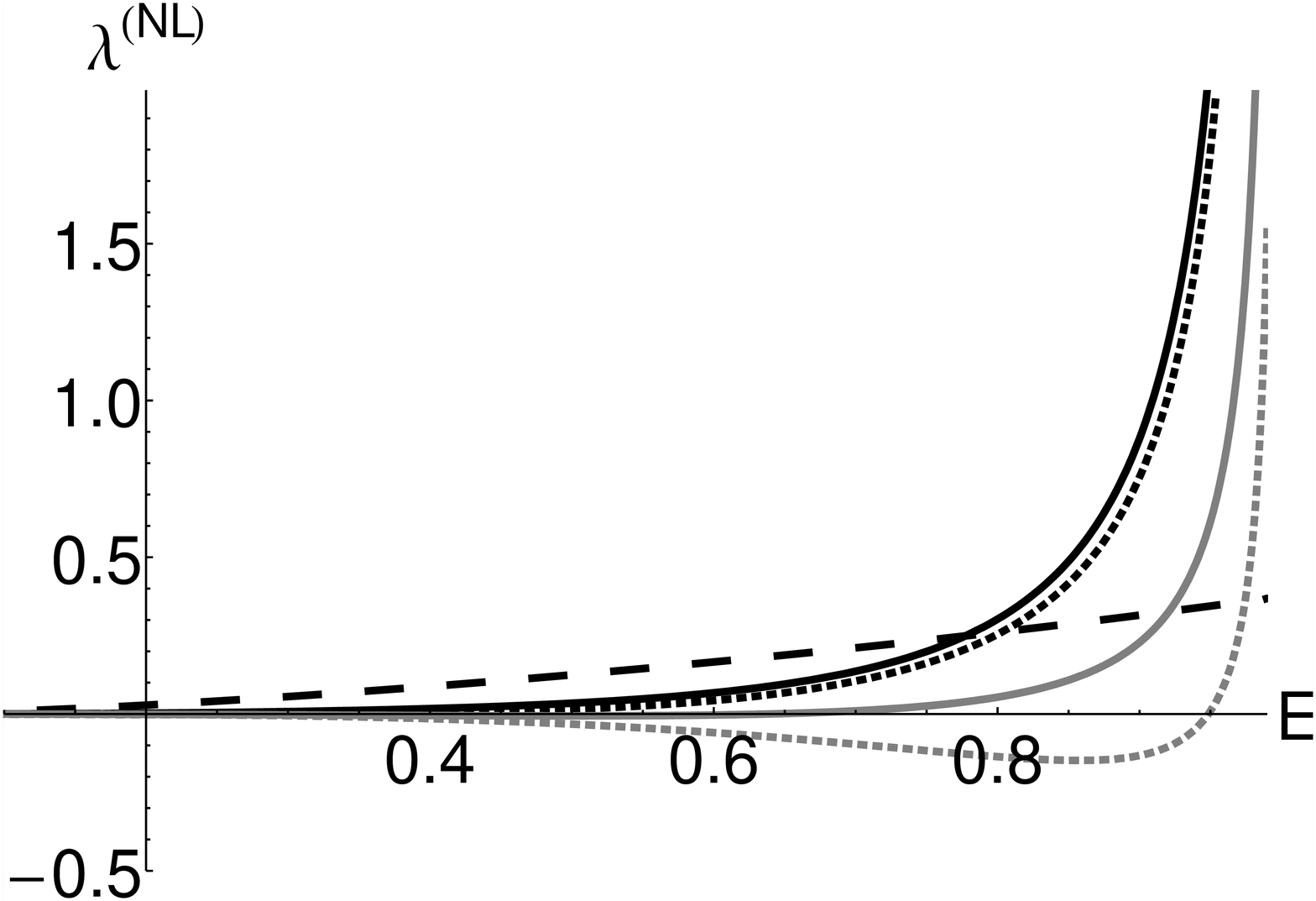} \includegraphics[width=0.48%
\columnwidth]{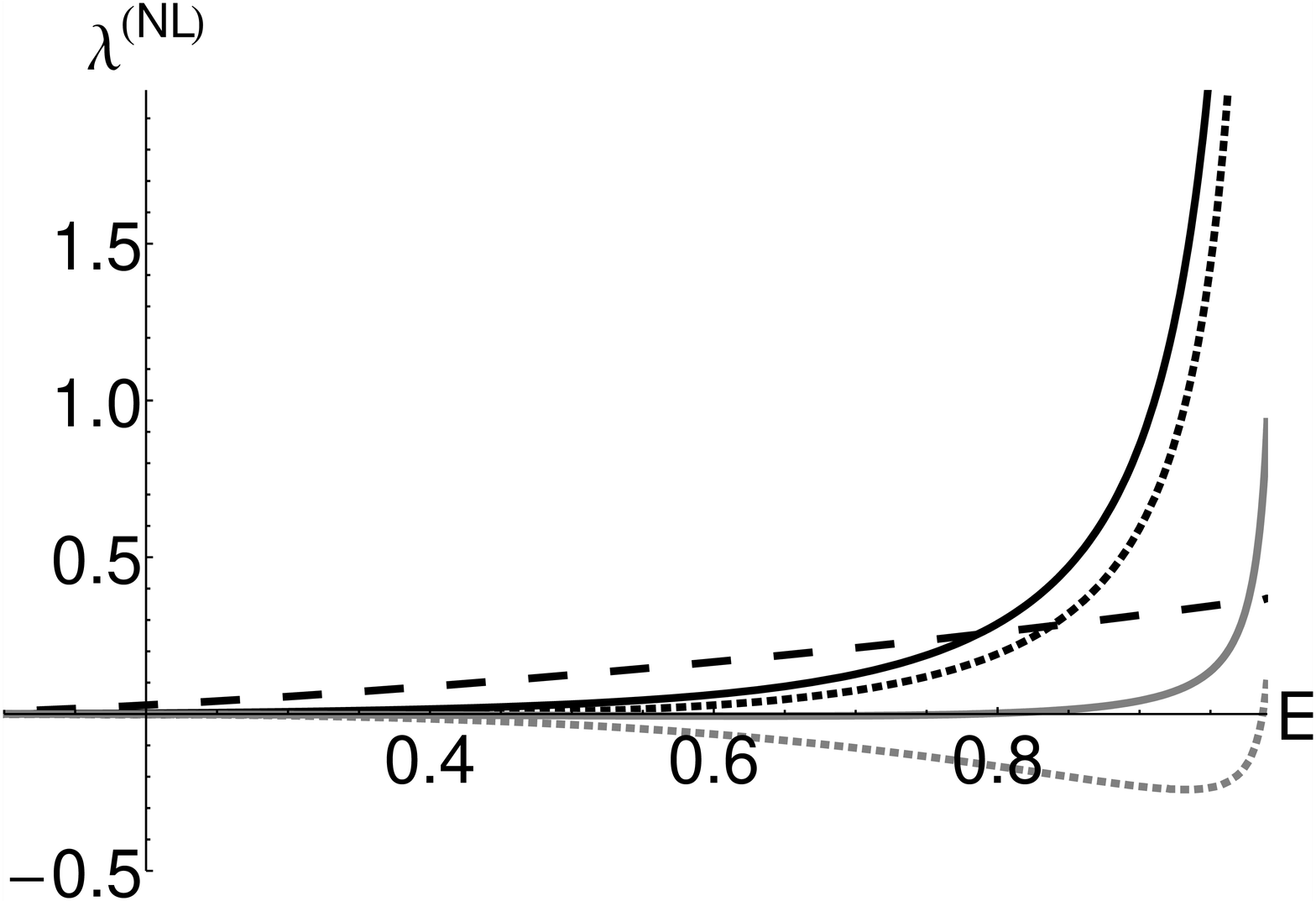}
\caption{Nonlinear contributions $\protect\lambda ^{\hbox{\tiny NL}}$ (see
Eqs. (\protect\ref{lambdas})) to intracavity squeezing, as functions of the
normalized pump amplitude $E$ for $\hat{\protect\kappa}_{0}=5$ (left) and $%
\hat{\protect\kappa}_{0}=10$ (right). In both plots: $\protect\lambda _{%
\protect\chi }^{\hbox{\tiny NL}}$ $\rightarrow $ black solid, $\protect%
\lambda _{\protect\mu }^{\hbox{\tiny NL}}\rightarrow $ black dashed, $%
\protect\lambda _{\protect\phi }^{\hbox{\tiny NL}}\rightarrow $ black
dotted, $\protect\lambda _{L}^{\hbox{\tiny NL}}\rightarrow $ gray solid, and 
$\protect\lambda _{T}^{\hbox{\tiny NL}}\rightarrow $ gray dotted.}
\label{f:fasi}
\end{figure}

For a balanced and exactly tuned OPO we have%
\begin{widetext}%
\begin{align}
\lambda _{\chi _{0}}^{NL}& =\frac{E\left( 2E^{3}\text{$\hat{\kappa}_{0}$}%
+2E^{2}\text{$\hat{\kappa}_{0}$}(2+\text{$\hat{\kappa}_{0}$})+(2+\text{$\hat{%
\kappa}_{0}$})^{2}-2E(-2+\text{$\hat{\kappa}_{0}$}(2+\text{$\hat{\kappa}_{0}$%
}))\right) }{2(1-E^{2})(1+E)(2+\text{$\hat{\kappa}_{0}$})(2+2E+\text{$\hat{%
\kappa}_{0}$})}~,  \notag \\
\lambda _{\phi _{P}}^{NL}& \simeq -\frac{\text{$\hat{\kappa}_{0}$}(2+\text{$%
\hat{\kappa}_{0}$})-E\text{$\hat{\kappa}_{0}$}(6+\text{$\hat{\kappa}_{0}$}%
)+2E^{3}(8+5\text{$\hat{\kappa}_{0}$})-2E^{2}(12+\text{$\hat{\kappa}_{0}$}(9+%
\text{$\hat{\kappa}_{0}$}))}{16\left( 1-E^{2}\right) \text{$\hat{\kappa}_{0}$%
}(2+\text{$\hat{\kappa}_{0}$})}~,  \notag \\
\lambda _{\nu }^{NL}& \simeq -\frac{E\left( E^{3}+2\text{$\hat{\kappa}_{0}$}%
-E^{2}(3+\text{$\hat{\kappa}_{0}$})-E(6+\text{$\hat{\kappa}_{0}$})\right) }{%
2(1-E^{2})(1+E)\text{$\hat{\kappa}_{0}$}^{2}}~,  \notag \\
\lambda _{T}^{NL}& \simeq \frac{E^{2}\left( 1+E\right) ^{2}-\left(
2+E\right) (1-E)\text{$\hat{\kappa}_{0}$}^{2}}{2(1-E^{2})\left( 1+E\right) 
\text{$\hat{\kappa}_{0}$}^{2}}~,  \notag \\
\lambda _{\mu _{P}}^{NL}& \simeq \frac{2+E^{2}}{2(1+E)^{2}}~.
\label{lambdas}
\end{align}%
\end{widetext}As it is apparent from the above formulas the contribution of $%
\hat{\mu}_{p}$ remains bounded on approaching the threshold ($E\rightarrow 1$%
), whereas the other terms diverge as $(1-E)^{-1}$, see Fig. \ref{f:fasi}.

It it noteworthy that the ratio between $\lambda _{\chi _{0}}^{%
\hbox{\tiny
NL}}$ and the analogous quantity $\lambda _{\chi _{0}}^{\hbox{\tiny PPSE}}$
calculated \ in \cite{Chaturvedi Dechoum Drummond} by means of the Positive
P representation (PPSE) goes as $\frac{\lambda _{\chi _{0}}^{\hbox{\tiny
NL}}}{\lambda _{\chi _{0}}^{\hbox{\tiny PPSE}}}\simeq \frac{\hat{\kappa}%
_{0}+2}{3\hat{\kappa}_{0}+2}+O(1-E)$. In the limiting case of $E\simeq 1$
the two approaches differ by a $\hat{\kappa}_{0}$-dependent factor bounded
between $1/3$ and $1$. Such a substantial agreement between the two results
in proximity of the singular point validates the use made in the present
paper of the extended GHLE.

\section{Kurtosis excess and comparison with the experimental results\label%
{s:extra}}

The field $\alpha _{out,1}$ outing the OPO is a function of $\alpha $, the
mirror damping coefficient $\gamma _{1}$, and the corresponding input noise $%
\mathbf{N}_{1}$\cite{Collett} 
\begin{equation}
\mathbf{\alpha }_{out,1}=g_{\chi }\sqrt{2\gamma _{1}}\left( \mathbf{\alpha }-%
\frac{1}{2\hat{\gamma}_{1}}\mathbf{N}_{1}\right) ~.  \label{fieldout}
\end{equation}%
Accordingly the generic output quadrature reads 
\begin{equation*}
X_{\theta }=\frac{1}{g_{\chi }^{2}\sqrt{2\gamma _{1}}}\mathbf{\theta }^{T}%
\mathbf{\cdot \alpha }_{out,1}=X_{\theta }^{\left( 1\right) }+X_{\theta
}^{\left( 2\right) }+\cdots
\end{equation*}%
where $\mathbf{\theta }=\frac{1}{2}\left( e^{-i\theta },e^{i\theta }\right) $
and $X_{\theta }^{\left( m\right) }$\ corresponds to $\alpha ^{\left(
m\right) }$. While $X_{\theta }^{\left( 1\right) }$ is Gaussian the terms $%
X_{\theta }^{\left( m>1\right) }$\ deviate from the normal distribution.

Quadratures are detected by a balanced homodyne and the relative current is
measured by selecting a frequency $\Omega _{f}$ and an integration time $1/%
\hat{\gamma}_{f}$ \cite{Wu,OPEX}. Accordingly, the detector output is
represented by%
\begin{equation}
V_{\theta }=\hat{F}_{f}X_{\theta }=\int_{-\infty }^{0}e^{\hat{\gamma}%
_{f}\tau ^{\prime }}\cos \left( \Omega _{f}\tau ^{\prime }\right) X_{\theta
}\left( \tau ^{\prime }\right) d\tau ^{\prime }  \label{filter}
\end{equation}%
The deviation of $V_{\theta }$ from a Gaussian distribution can be measured
by the \textit{Kurtosis--excess} parameter $K_{\theta }$ (see Appendix \ref%
{Kexpansion} for details)%
\begin{equation}
K_{\theta }=\frac{\left\langle :V_{\theta }^{4}:\right\rangle -3\left\langle
:V_{\theta }^{2}:\right\rangle ^{2}}{3\left\langle :V_{\theta
}^{2}:\right\rangle ^{2}}\simeq \sum_{\iota }g_{\iota }^{2}\frac{\Upsilon
_{\theta \iota }}{\left\langle :V_{\theta }^{\left( 1\right)
2}:\right\rangle _{\chi }^{2}}  \label{Kurtosi}
\end{equation}%
with $\Upsilon _{\theta \iota }$ given in Eq. (\ref{integral}), the weight
of the different noise sources indicated generally by $N_{\iota }$.

The spectral density $\tilde{S}_{\mu _{p}}\left( w\right) =\left\langle 
\tilde{N}_{\mu _{p}}\left( -w\right) \tilde{N}_{\mu _{p}}\left( w\right)
\right\rangle $ extends generally up to $1\div 2$ $MHz$. For the sake of
simplicity it has been approximated by a uniform spectrum extending up to 1
MHz . $\delta \tilde{\nu}$ and $\delta \tilde{T}$ extend on very narrow
bandwidths, while $\tilde{N}_{\chi _{0}}$ and $\tilde{\varpi}_{p}$ are white
noise sources.

\begin{figure}[h]
\includegraphics[width=0.95\columnwidth]{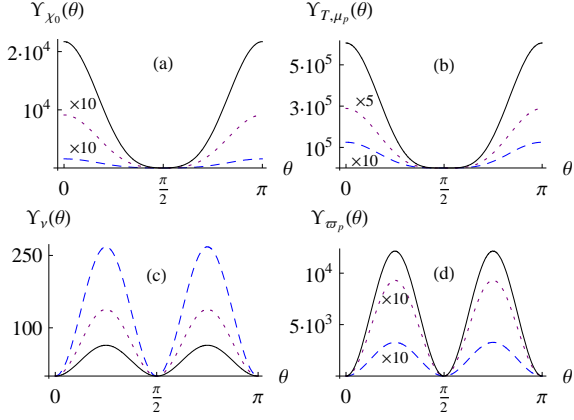}
\caption{$\Upsilon _{\protect\theta \protect\iota }$ (see Eq. (\protect\ref%
{integral})) vs. $\protect\theta $ for different noise sources: (a) pump
quantum noise; (b) amplitude and temperature; (c) cavity length; (d) pump
phase. The plots, referred to different scaling factors, have been
calculated for typical values of $\Omega _{f}~(=0.3)$, $\protect\gamma %
_{f}~(=0.15)$ and $E=0.71($dashed blue), $0.87$ (purple dotted), $0.975$
(black).}
\label{K}
\end{figure}

In Fig. \ref{K}(a--d) we have plotted the five $\Upsilon _{\iota }\left(
\theta \right) $ vs. $\theta $ for three excitation strengths $%
E=0.71,0.87,0.975$ of a perfectly tuned OPO with a pump cavity mode
linewidth twice the signal one ($\hat{\kappa}_{0}=2$), a condition similar
to that of Ref.\cite{OPEX}. The graphs show that for $N_{\chi _{0}}$, $\hat{%
\mu}_{p}$, and $\delta \hat{T}$ the maximum deviation from a Gaussian
appears for $\theta =0$, while for $\varpi _{p}$ and $\delta \hat{\nu}$ it
occurs for $\theta =\pm \pi /4$. (see Eq. (\ref{B1i})).

\begin{figure}[h]
\includegraphics[width=0.95\columnwidth]{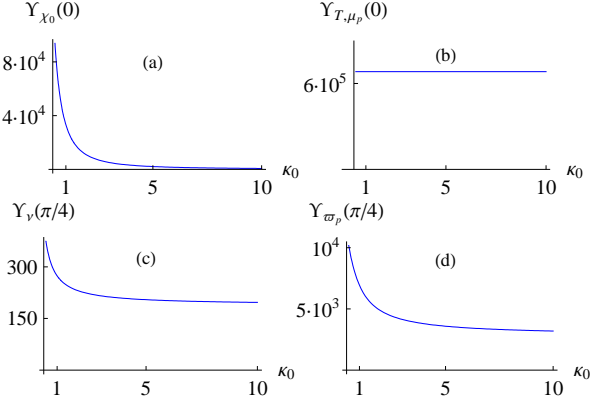}
\caption{Maxima of $\Upsilon _{\protect\theta \protect\iota }$ (see Eq. (%
\protect\ref{integral}) vs. $\protect\kappa _{0}$ for the conditions of Fig. 
\protect\ref{K}. ($E\rightarrow 0.975$).}
\label{kappa0}
\end{figure}
The maxima of $\Upsilon _{\iota }\left( \theta \right) $ for $N_{\chi _{0}}$%
, $\varpi _{p}$ and $\delta \hat{\nu}$ decrease for increasing $\hat{\kappa}%
_{0}$ (see Fig. \ref{kappa0}(a--d)). As expected from Eq. (\ref{B1}), the
contribution of $\delta \hat{T}$ is independent of $\hat{\kappa}_{0}$. The
same holds true approximately for $\hat{\mu}_{p}$ too having considered a
technical noise bandwidth small compared to $\hat{\kappa}_{0}.$

Looking at Fig. \ref{kappa0} we see that the maximum for $\Upsilon _{\mu
_{p},T}$ is at least one order of magnitude larger than the other ones.
Moreover, $g_{\mu _{p}}\gg g_{T}$ so that for pump level up to $E^{2}=0.95$,
the NG behavior is essentially due to the laser amplitude fluctuations. 
\begin{figure}[h]
\includegraphics[width=0.95\columnwidth]{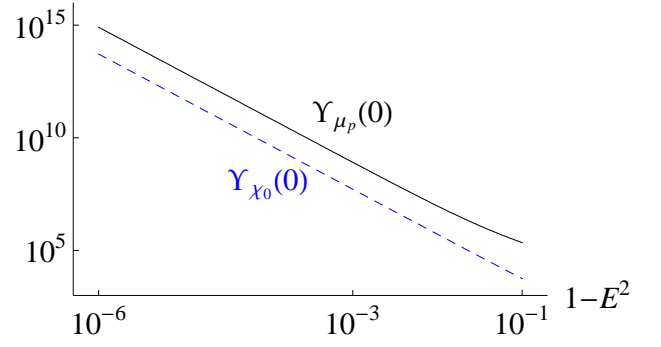}
\caption{$\Upsilon _{0\protect\chi _{0}}$ (pump quantum noise) and $\Upsilon
_{0\protect\mu _{p}}$ (amplitude fluctuations) vs. $1-E^{2}$.}
\label{EE}
\end{figure}

In Fig. \ref{EE} $\Upsilon _{0\chi _{0}}$ (blue--dashed) and $\Upsilon
_{0\mu _{p}}$ (black--solid) are plotted vs. $1-E^{2}$ on a double
logarithmic scale. Approaching the threshold the influence of $N_{\chi _{0}}$
increases dramatically although $\Upsilon _{0\chi _{0}}$ does not overcome $%
\Upsilon _{0\mu _{p}}$. Moreover, $g_{\chi }\ll g_{\mu _{p}}$, so that the
observation of pure quantum effects, predicted by Drummond \textit{et al.} 
\cite{Chaturvedi Dechoum Drummond}, is demanded to future technology when
either new materials with huge non--linear coefficients (enhanced $g_{\chi }$%
) or very quiet lasers (reduced $g_{\mu _{p}}$) will be available.

\begin{figure}[h]
\includegraphics[width=0.95\columnwidth]{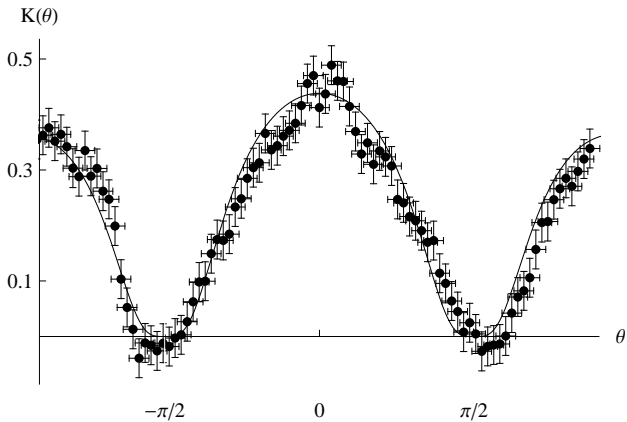}
\caption{$K_{\protect\mu _{p}}\left( \protect\theta \right) $ (see Eq. (%
\protect\ref{Kurtosi})) overimposed to the experimental data of Ref. 
\protect\cite{OPEX}. $K_{\protect\mu _{p}}\left( \protect\theta \right) $
has been calculated by assuming $E^{2}=0.92$, $\Omega _{f}=0.3$, $\protect%
\gamma _{f}=0.15$ and $g_{\protect\mu _{P}}=0.007$. \ The horizontal error
bar accounts for the detector phase $\protect\theta $ stability, while the
vertical one corresponds to the average spread between two neighbour $K_{%
\protect\theta }$ data.}
\label{exptoth}
\end{figure}

\begin{figure}[h]
\includegraphics[width=0.95\columnwidth]{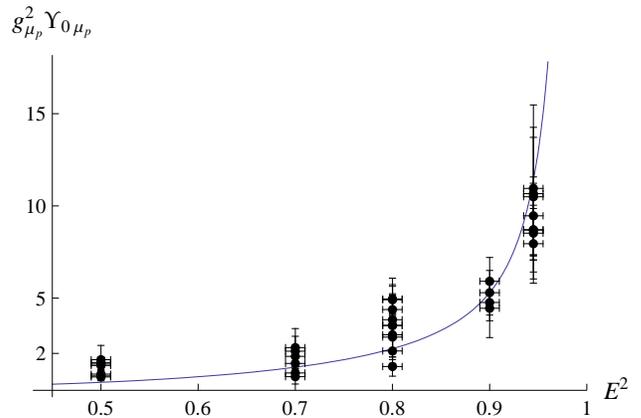}
\caption{Comparison of $g_{\protect\mu _{p}}^{2}\Upsilon _{0\protect\mu %
_{p}} $ (Eq. (\protect\ref{Kurtosi})) vs. $E^{2}$ (solid curve; $\Omega
_{f}=0.3$, $\protect\gamma _{f}=0.15$ and $g_{\protect\mu _{P}}=0.007$) with
experimental data. Each experimental point has been obtained multiplying the
measured kurtosis at $\protect\theta =0$ for the relative squared variance.
It is evident the agreement between data and theoretical curve within error
bars.}
\label{maxvsE}
\end{figure}

Being the laser amplitude noise the prominent source influencing the
non--linear behavior we have compared some experimental findings of Ref. 
\cite{OPEX} with the predictions of the herein discussed model. In
particular, the experimental behaviors of $K_{\theta }$ and $K_{\theta \mu
_{p}}$ vs. $\theta \in (-\pi ,\pi )$ are reported in Fig. \ref{exptoth}.
Moreover, the maximum value of the experimental kurtosis, for $\theta =0$,
and $g_{\mu _{p}}^{2}\Upsilon _{0\mu _{p}}$ (see Eq. (\ref{Kurtosi})) vs. $%
E^{2}$ $\in (0.45,0.97)$ are plotted in Fig. \ref{maxvsE}.

In general $\Upsilon _{\theta \mu _{p}}$ can be represented by 
\begin{equation}
\Upsilon _{\theta \mu _{p}}=\Upsilon _{4\mu _{p}}\cos 4\theta +\Upsilon
_{2\mu _{p}}\cos 2\theta +\Upsilon _{0\mu _{p}}  \label{Ymu}
\end{equation}%
with $\Upsilon _{4,2,0\mu _{p}}$ functions of $E,\hat{\kappa}_{0}$ and $\hat{%
\kappa}$. For very small deviations from the resonant configuration and $%
\left\vert E\right\vert $ close to $1$, $\Upsilon _{\theta \mu _{p}}$
depends critically on $\left\vert E\right\vert $ 
\begin{equation}
\left\vert E\right\vert =\frac{E_{0}}{\sqrt{\left( 1+4\frac{\nu _{p}^{2}}{%
\gamma _{s}^{2}}\right) \left( 1+\frac{\nu _{p}^{2}}{\gamma _{p}^{2}}\right) 
}}  \label{excitation}
\end{equation}%
with $E_{0}=\epsilon |\bar{\chi}|/\left( \gamma _{p}\gamma _{s}^{3/2}\right) 
$ the excitation strength at resonance.

The different heights of the peaks in the experimental data (see Fig. \ref%
{exptoth}) can be can be ascribed to the variation of $E$ during a
measurement. The acquisition time for $V_{\theta }$ for each $\theta $
lasted about 2 ms, implying a total $\theta -$scanning acquisition time of
200 ms. The apparatus was equipped with a digital controller providing a
crystal temperature time constant $>10^{3}$ s and a Drever-Pound system
controlling the cavity tuning with a time constant $>10$ s. However, during
the total acquisition time, slow drifts of the average cavity detuning $\nu
_{p}$ can occur thus inducing a variation of the effective excitation
parameter $E$ during the scan (see Eq. (\ref{excitation})). The simplest and
more direct way to account for the time dependence of $\nu _{p}$ is to set
in (\ref{excitation}) 
\begin{equation}
\nu _{p}=\alpha \left( \theta -\theta _{0}\right)  \label{detuning}
\end{equation}%
with $\alpha $ and $\theta _{0}$ two fitting parameters and express the
coefficients $\Upsilon _{4,2,0\mu _{p}}$ in Eq. (\ref{Ymu}) as functions of $%
\theta $.

The best fit (continuous curve of Fig. \ref{exptoth}) of the experimental
data has been computed by assuming a spectral density $\tilde{S}_{\mu _{p}}$
uniform in the interval $0\div 1.0$ $MHz$, in agreement with the laser
(Lightwave mod. 142) technical noise specification and optimizing $E_{0}$
(see (\ref{excitation})), $\alpha $ and $\theta _{0}$ (see (\ref{detuning}%
)). The best agreement has been obtained for $\alpha =0.013$, $\theta
_{0}=\pi $ and $E_{0}=0.932$, values corresponding to a drift of the
resonance frequency of $\approx 4\%$ the pump cavity mode linewidth and a
variation of $\left\vert E\right\vert $ of 0.006. Finally, normalizing $%
g_{\mu _{p}}^{2}\Upsilon _{\theta \mu _{p}}$ to the squared experimental
variance (see Eq. (\ref{Kurtosi})) the best agreement was obtained for $%
g_{\mu _{p}}=0.007$ in agreement with laser noise specification ($<1\%$).

The NG character depends critically on the distance from the threshold (see
Fig. \ref{EE}). $g_{\mu _{p}}^{2}\Upsilon _{0\mu _{p}}$ (see Eq. (\ref%
{Kurtosi})) vs. $E^{2}$ is compared to a set of 35 data obtained for five
different values of $E^{2}$ ($0.5$, $0.7$, $0.8$, $0.9$, and $0.95$) in Fig. %
\ref{maxvsE}. Plotted values have been obtained multiplying the experimental
kurtosis by the relative squared variance. Error bars have been obtained by
standard error propagation. The good agreement between the expected behavior
and the data confirms, once more, the effectiveness of the model and the
validity of assumptions about the relative noise strengths.

\section{Conclusions\label{s:outro}}

The statistical properties of the fields generated by an OPO depend on the
fluctuations of many classical parameters, namely pump amplitude ($g_{\mu
_{p}}\hat{\mu}_{p}$), pump frequency ($g_{\varpi _{p}}\varpi _{p}$), cavity
detuning ($g_{\nu }\delta \hat{\nu}$), and crystal temperature ($g_{T}\delta 
\hat{T}$). In this paper it has been presented a model of the OPO based on
an extension of the Graham-Haken quantum Langevin equations (GHLE) which
accounts for the fluctuations of these parameters. The field generated has
been dealt with as the response of a nonlinear device to these noise
sources. Then, expanding the extended GHLE system at different orders in $%
g_{\mu _{p}},g_{\varpi _{p}},g_{\nu },g_{T}$ a hierarchy of equations has
been obtained with $\hat{\mu}_{p},\varpi _{p},\delta \hat{\nu},\delta \hat{T}
$ acting as noise sources together with the quantum noises ($g_{\chi }\hat{N}%
_{\chi },g_{\chi }\hat{N}_{\chi _{0}}$) entering the optical cavity. These
sources have been modeled as Gaussian processes with unit standard
deviations weighted by the respective factors $g_{\mu _{p}},g_{\varpi
_{p}},g_{\nu },g_{T}$ typically ranging in the intervals $10^{-2}\div
10^{-1},10^{-7/2}\div 10^{-2},10^{-5}\div 10^{-1},10^{-5}\div 10^{-4}$.

The extended GHLE solutions, obtained beyond the linear approximation, have
been used for assessing the non-Gaussian character of the field outing a
degenerate OPO. The departure of the output quadrature $X_{\theta }$ from
the Gaussian statistics has been estimated by means of the \textit{%
Kurtosis--excess }figure $K_{\theta }=\left( \left\langle X_{\theta
}^{4}\right\rangle -3\left\langle X_{\theta }^{2}\right\rangle ^{2}\right)
/\left\langle X_{\theta }^{2}\right\rangle ^{2}$, i.e. the relative
deviation of $X_{\theta }$ 4-th moment from the Gaussian expression of it.
The model furnishes $K_{\theta }$ as a sum of contributions from different
parameters. Sets of plots have been provided, showing the dependence of $%
K_{\theta }$ on the OPO operating condition, namely, the ratio pump/signal
bandwidths, the excitation strength $E$, and the detection frequency $\Omega
_{f}$ and bandwidth $\gamma _{f}$. For typical operating conditions the pump
technical noise emerges as the most critical factor.

The model has been used for reproducing the experimental values, reported in
Ref. \cite{OPEX}, of $K_{\theta }$ vs. $\theta $ and $K_{0}$ vs. $E^{2}$.
The good agreement, within the error bars, of the experimental data with the
analythic predictions confirms the validity of the presented model. By
providing a physically-insightful and computationally-effective
parameterization of the OPO, the model may help in addressing the generation
of non--Gaussian states by means of OPO sources.

\appendix

\section{Linearization below threshold}

The Fourier transform of the Green's functions $\mathbf{G,G}_{0}$ relative
to a degenerate OPO are given by 
\begin{equation}
\mathbf{\tilde{G}}=\frac{1}{\tilde{D}}\left[ 
\begin{array}{cc}
\tilde{\Delta}^{\ddagger } & e^{-i\vartheta }\left\vert E\right\vert \\ 
e^{i\vartheta }\left\vert E\right\vert & \tilde{\Delta}%
\end{array}%
\right] \;,\;\mathbf{\tilde{G}}_{0}=\left[ 
\begin{array}{cc}
\tilde{\Delta}_{0}^{-1} & 0 \\ 
0 & \tilde{\Delta}_{0}^{\ddagger -1}%
\end{array}%
\right] \;  \label{components matrix Gopo bt}
\end{equation}%
with$~\vartheta =\psi -\frac{1}{2}\psi _{0}$, $\tilde{\Delta}=\hat{\kappa}%
-i\omega $, $\tilde{D}\left( \omega \right) =\tilde{\Delta}\tilde{\Delta}%
^{\ddagger }-\left\vert E\right\vert ^{2}=-\left( \omega +\omega _{+}\right)
\left( \omega +\omega _{-}\right) $, $\tilde{\Delta}_{0}=\hat{\kappa}%
_{0}-i\omega ,\omega _{\pm }=i\left( \cos \psi \mp \sqrt{\left\vert
E\right\vert ^{2}-\sin ^{2}\psi }\right) .$ In particular for $\hat{\kappa}%
=1 $ (tuned device) they correspond in the time domain to 
\begin{align}
\mathbf{G}\left( \tau \right) & =\left\{ 
\begin{array}{c}
e^{-\tau }\left[ 
\begin{array}{cc}
\cosh \left( E\tau \right) & \sinh \left( E\tau \right) \\ 
\sinh \left( E\tau \right) & \cosh \left( E\tau \right)%
\end{array}%
\right] \;\tau >0 \\ 
0\;\;\;\;\;\;\;\;\;\;\;\;\;\;\;\;\;\;\;\;\;\;\;\;\;\;\;\;\;\;\ \tau <0%
\end{array}%
\right.  \notag \\
\mathbf{G}_{0}\left( \tau \right) & =\left\{ 
\begin{array}{c}
e^{-\hat{\kappa}_{0}\tau }\left[ 
\begin{array}{cc}
1 & 0 \\ 
0 & 1%
\end{array}%
\right] \;\tau >0 \\ 
0\;\;\;\;\;\;\;\;\;\;\;\;\;\;\;\;\tau <0%
\end{array}%
\right.  \label{Green time}
\end{align}%
%
%
%
%
%
%
%
%
%
%
%
%
%
%
%
%
%
%
%
%
%
%
%
%
%
%
%
%
%
%
%
%
%
%
%
%
%
%
%
%
%
%
%
%
%
%
%
%
%
%
%
%
%
%
%
%
%
%
%
%
%
%
%
%
%
%
%
%
%
%
%
%
%
%
%
%
%
%
%
%
%
%
%
%
%
%
%
%
%
%
%
%
%
%
%
%
%
%
%
%
%
%
%
%
%
%
%
%
%
%
%
%
%
%
%
%
%
%
%
%
%
%
%
%
%
%
%
%
%
%
%
%
%
%
%
%
%
%
%
%
%
%
%
%
%
%
%
%
%
%
%
%
%
%%%%%%%%%%%%%%%%%%%%%%%%%%%%%%

\section{Time-normal ordered correlation matrix $:\mathbf{\tilde{\protect%
\sigma}}^{\left( 1,1\right) }\left( \protect\omega \right) :$}

The Fourier transform $:\mathbf{\tilde{\sigma}}^{\left( 1,1\right) }\left(
\omega \right) :$ of the time-normal ordered matrix $:\boldsymbol{\alpha }%
\left( \tau -\tau ^{\prime }\right) :=:\left\langle \boldsymbol{\alpha }%
^{\left( 1\right) }\left( \tau \right) \mathbf{\alpha }^{\left( 1\right)
T}\left( \tau ^{\prime }\right) \right\rangle _{\chi }:$ ($:\boldsymbol{%
\alpha }\left( \tau ^{\prime }-\tau \right) :=:\boldsymbol{\alpha }\left(
\tau -\tau ^{\prime }\right) :$ and $:\boldsymbol{\alpha }_{a^{\dagger
}a^{\dagger }}:=:\boldsymbol{\alpha }_{aa}^{\ast }:$, $:\boldsymbol{\alpha }%
_{aa^{\dagger }}:=:\boldsymbol{\alpha }_{a^{\dagger }a}:$) is obtained from $%
\mathbf{\tilde{\sigma}}^{\left( 1,1\right) }\left( \omega \right) =\mathbf{%
\tilde{\sigma}}\left( \omega \right) /\left[ \left( \omega ^{2}-\omega
_{+}^{2}\right) \left( \omega ^{2}-\omega _{-}^{2}\right) \right] $ with%
\begin{equation*}
\mathbf{\tilde{\sigma}}\left( \omega \right) =4\left[ 
\begin{array}{cc}
\frac{\tilde{\Delta}^{\ddagger }}{e^{-i\vartheta }\left\vert E\right\vert }
& \frac{\tilde{\Delta}^{\ddagger }\left( -\omega \right) \tilde{\Delta}%
\left( \omega \right) }{\left\vert E\right\vert ^{2}} \\ 
1 & \frac{\tilde{\Delta}}{e^{i\vartheta }\left\vert E\right\vert }%
\end{array}%
\right]
\end{equation*}%
by first normally ordering $\mathbf{\tilde{\sigma}}\left( \omega \right) ,$ 
\begin{equation*}
\mathbf{\tilde{\sigma}}\left( \omega \right) \rightarrow \boldsymbol{\tilde{%
\sigma}}_{N}\left( \omega \right) =4\left[ 
\begin{array}{cc}
\frac{\tilde{\Delta}^{\ddagger }}{e^{-i\vartheta }\left\vert E\right\vert }
& 1 \\ 
1 & \frac{\left( \tilde{\Delta}^{\ddagger }\right) ^{\ast }}{e^{i\vartheta
}\left\vert E\right\vert }%
\end{array}%
\right]
\end{equation*}%
and then, symmetrizing with respect to time reversal%
\begin{eqnarray}
&:&\mathbf{\tilde{\sigma}}^{\left( 1,1\right) }\left( -\omega \right) :=-%
\frac{2\omega _{+}}{\omega ^{2}-\omega _{+}^{2}}\boldsymbol{\tilde{\sigma}}%
_{N}\left( -\omega _{+}\right) -\frac{2\omega _{-}}{\omega ^{2}-\omega
_{-}^{2}}\boldsymbol{\tilde{\sigma}}_{N}\left( -\omega _{-}\right)  \notag \\
&=&\frac{\mathbf{\tilde{\sigma}}_{TN}\left( \omega \right) }{\left( \omega
^{2}-\omega _{+}^{2}\right) \left( \omega ^{2}-\omega _{-}^{2}\right) }
\label{time-normal ordered}
\end{eqnarray}%
In particular for the tuned case 
\begin{equation}
\mathbf{\tilde{\sigma}}_{TN}\left( \omega \right) =4\left[ 
\begin{array}{cc}
\frac{1+\text{$E$}^{2}+\omega ^{2}}{2E} & 1 \\ 
1 & \frac{1+\text{$E$}^{2}+\omega ^{2}}{2E}%
\end{array}%
\right]  \label{time-normal ordered reduced}
\end{equation}

%
%
%
%
%
%
%
%
%
%
%
%
%
%
%
%
%
%
%
%
%
%
%
%
%
%
%
%
%
%
%
%
%
%
%
%
%
%
%
%
%
%
%
%
%
%
%
%
%
%
%
%
%%%%%%%%%%%%%%%%%%%%%%%%%

\section{$\mathbf{B}^{\left( 1\right) }$ expansion}

The different noise sources $N_{\iota }$ contribute to $\mathbf{\tilde{B}}%
^{\left( 1\right) }$ (Eq. (\ref{B1})) through the terms: 
\begin{equation}
\mathbf{\tilde{B}}^{\left( 1\right) }=g_{\chi }\tilde{N}_{\chi _{0}}\mathbf{%
\tilde{B}}_{\chi }^{\left( 1\right) }+g_{\chi }\tilde{N}_{\chi
_{0}}^{\ddagger }\mathbf{\tilde{B}}_{\chi }^{\left( 1\right) T}+\sum_{i}g_{i}%
\tilde{N}_{i}\mathbf{\tilde{B}}_{i}^{\left( 1\right) }\;,  \label{B1a}
\end{equation}%
with%
\begin{align}
\mathbf{\tilde{B}}_{\chi }^{\left( 1\right) }& =E\left[ 
\begin{array}{ll}
0 & \tilde{\Delta}_{0}^{-1} \\ 
0 & 0%
\end{array}%
\right]  \notag \\
\;\mathbf{\tilde{B}}_{\mu _{_{P}}}^{\left( 1\right) }& =\left[ 
\begin{array}{ll}
0 & e^{-i\vartheta }\left\vert E\right\vert \hat{\kappa}_{0}\tilde{\Delta}%
_{0}^{-1} \\ 
e^{i\vartheta }\left\vert E\right\vert \hat{\kappa}_{0}^{\ast }\tilde{\Delta}%
_{0}^{\ddagger -1} & 0%
\end{array}%
\right]  \notag \\
\mathbf{\tilde{B}}_{\varpi _{P}}^{\left( 1\right) }& =i\left[ 
\begin{array}{ll}
\frac{1}{2} & e^{-i\vartheta }\left\vert E\right\vert \tilde{\Delta}_{0}^{-1}
\\ 
-e^{i\vartheta }\left\vert E\right\vert \tilde{\Delta}_{0}^{\ddagger -1} & -%
\frac{1}{2}%
\end{array}%
\right]  \notag \\
~\mathbf{\tilde{B}}_{T}^{\left( 1\right) }& =\left[ 
\begin{array}{ll}
0 & e^{-i\vartheta }\left\vert E\right\vert \\ 
e^{i\vartheta }\left\vert E\right\vert & 0%
\end{array}%
\right]  \notag \\
\mathbf{\tilde{B}}_{\nu }^{\left( 1\right) }& =i\left[ 
\begin{array}{ll}
-1 & e^{-i\vartheta }\left\vert E\right\vert \tilde{\Delta}_{0}^{-1} \\ 
-e^{i\vartheta }\left\vert E\right\vert \tilde{\Delta}_{0}^{\ddagger -1} & 1%
\end{array}%
\right]  \label{B1i}
\end{align}%
%
%
%
%
%
%
%
%
%
%
%
%
%
%
%
%
%
%
%
%
%
%
%
%
%
%
%
%
%
%
%
%
%
%
%
%
%
%
%
%
%
%
%
%
%
%
%
%
%
%
%
%
%
%
%
%
%
%
%
%
%
%
%
%
%
%
%
%
%
%
%
%
%
%
%
%
%
%
%
%
%
%
%
%
%
%
%
%
%
%
%
%
%
%
%
%
%
%
%
%
%
%
%
%
%
%
%
%
%
%
%
%
%
%
%
%
%
%
%
%
%
%
%
%
%
%
%
%
%
%
%
%
%
%
%
%
%
%
%
%
%%%%%%%%%%%%%%%%%%%%

\section{Kurtosis--excess expansion\label{Kexpansion}}

From the vanishing of the time-normal ordered correlations $\left\langle 
\mathbf{:\alpha }^{\left( 1\right) }\mathbf{N}_{1}^{T}:\right\rangle
=\left\langle \mathbf{:\alpha }^{\left( 2\right) }\mathbf{N}%
_{1}^{T}:\right\rangle =0$, it follows that:%
\begin{equation*}
\left\langle :X_{\theta }^{\left( l\right) }X_{\theta }^{\left( m\right)
}:\right\rangle _{\chi }=\mathbf{\theta }^{T}\mathbf{\cdot }\left\langle 
\mathbf{:\alpha }^{\left( l\right) }\mathbf{\alpha }^{\left( m\right)
T}:\right\rangle _{\chi }\mathbf{\cdot \theta }
\end{equation*}%
with $l,m=1,2$. Hence, retaining only the lowest non linear orders\textbf{\ }%
for\textbf{\ }$\left\langle :V_{\theta }^{4}:\right\rangle $ and $%
\left\langle :V_{\theta }^{2}:\right\rangle ^{2}$, $\Upsilon _{\theta \iota }
$ (see Eq. (\ref{Kurtosi})) reads:%
\begin{eqnarray}
\Upsilon _{\theta \iota } &=&\left\langle \left( \left\langle :V_{\theta
}^{\left( 1\right) }V_{\theta \iota }^{\left( 2\right) }+V_{\theta \iota
}^{\left( 2\right) }V_{\theta }^{\left( 1\right) }:\right\rangle _{\chi
}\right) ^{2}\right\rangle _{\iota }  \notag \\
&=&\frac{1}{2\pi }\int_{-\infty }^{\infty }\tilde{S}_{\iota }\left( w\right) 
\mathbf{\theta }^{T}\cdot \mathbf{\tilde{\varsigma}}_{\iota }\left(
-w\right) \cdot \boldsymbol{\theta \theta }^{T}\boldsymbol{\cdot }\mathbf{%
\tilde{\varsigma}}_{\iota }\left( w\right) \cdot \mathbf{\theta }dw  \notag
\\
&=&\Upsilon _{4\iota }\cos 4\theta +\Upsilon _{2\iota }\cos 2\theta
+\Upsilon _{0\iota }  \label{integral}
\end{eqnarray}%
with $\tilde{S}_{\iota }\left( w\right) =\left\langle \tilde{N}_{i}\left(
-w\right) \tilde{N}_{i}\left( w\right) \right\rangle $ and $\tilde{\varsigma}%
_{\iota }$ a $2\times 2$ matrix%
\begin{eqnarray}
\tilde{\varsigma}_{\iota }\left( w\right)  &=&\left( 1+\hat{T}\right)
\left\langle :\left( \hat{F}_{f}\mathbf{\alpha }_{\iota }^{\left( 2\right)
}\left( 0\right) \right) \hat{F}_{f}\mathbf{\alpha }^{\left( 1\right)
T}\left( 0\right) :\right\rangle _{\chi }  \notag \\
&=&\frac{1}{2\pi }\frac{1}{\hat{\kappa}_{0}-iw}\int\nolimits_{-\infty
}^{\infty }\tilde{H}\left( w,\omega \right) \boldsymbol{\tilde{\sigma}}%
_{\iota }\left( w,\omega \right) d\omega \;,  \label{V1V2}
\end{eqnarray}%
$\hat{T}$ being a matrix transposition operator, $\boldsymbol{\tilde{\sigma}}%
_{\iota }\left( w,\omega \right) $ an entire function of $\omega $ and $w$%
\begin{multline*}
\boldsymbol{\tilde{\sigma}}_{\iota }\left( w,\omega \right) =\left( 1+\hat{T}%
\right) \left( \hat{\kappa}_{0}-iw\right) \tilde{D}\left( w+\omega \right) 
\mathbf{\tilde{G}}\left( w+\omega \right)  \\
\cdot \mathbf{\tilde{B}}_{\iota }^{\left( 1\right) }\left( w\right) \cdot 
\mathbf{\tilde{\sigma}}_{TN}\left( \omega \right) 
\end{multline*}%
with $\mathbf{\tilde{\sigma}}_{TN}\left( \omega \right) $ defined in Eq. (%
\ref{time-normal ordered}), 
\begin{equation*}
\tilde{H}\left( w,\omega \right) =\frac{\tilde{F}_{f}\left( \omega +w\right) 
\tilde{F}_{f}\left( -\omega \right) }{\tilde{D}\left( w+\omega \right)
\left( \omega ^{2}-\omega _{+}^{2}\right) \left( \omega ^{2}-\omega
_{-}^{2}\right) }
\end{equation*}%
and 
\begin{equation*}
\tilde{F}_{f}\left( \omega \right) =\frac{i}{2}\left( \frac{1}{\omega
-\Omega _{-}}+\frac{1}{\omega -\Omega _{+}}\right) 
\end{equation*}%
the Fourier transform of $\hat{F}_{f}$ (see Eq. (\ref{filter})) while $%
\Omega _{\pm }=\pm \Omega _{f}-i\gamma _{f}.$

Since $\lim\limits_{\left\vert \omega \right\vert \rightarrow \infty }\omega 
\tilde{H}\left( w,\omega \right) \boldsymbol{\tilde{\sigma}}_{\iota }\left(
w,\omega \right) =0$ the RHS of Eq. (\ref{V1V2}) is given by the sum of
residues%
\begin{equation}
\tilde{\varsigma}_{\iota }\left( w\right) =\frac{i}{\hat{\kappa}_{0}-iw}%
\sum_{l=1}^{4}H^{\left( l\right) }\left( w\right) \boldsymbol{\tilde{\sigma}}%
_{\iota }\left( w,\omega _{l}\right)  \label{sigmai}
\end{equation}%
where $H^{\left( l\right) }\left( w\right) =\mathtt{Res}_{\omega =\omega
_{l}}[H\left( w,\omega \right) ]$ for $\omega _{l}=\omega _{+},\omega
_{-},-\Omega _{+}$,$-\Omega _{-}$ $\left( l=1,2,3,4\right) $, poles of $%
\tilde{H}\left( w,\omega \right) $ in the upper complex $\omega $--plane.

In the limiting case of zero centered delta--like sources $\tilde{N}_{i}$
(see Eq. (\ref{B1a}))%
\begin{equation*}
\Upsilon _{\theta i}=\mathbf{\theta }^{T}\cdot \mathbf{\tilde{\varsigma}}%
_{i}\left( 0\right) \cdot \boldsymbol{\theta \theta }^{T}\boldsymbol{\cdot }%
\mathbf{\tilde{\varsigma}}_{i}\left( 0\right) \cdot \mathbf{\theta }
\end{equation*}%
Such an approximation holds true for $\tilde{N}_{T},\tilde{N}_{\nu }$ and in
a less measure for $\tilde{N}_{\mu _{p}},$ depending on the laser technical
noise bandwidth normalized to the OPO cavity one. On the other extreme, $%
\tilde{N}_{\chi _{0}}$ and $\tilde{N}_{\varpi _{p}}\ $represent white noises
processes for which $\Upsilon _{\theta \chi _{0},\varpi _{p}}$ reduce to%
\begin{multline*}
\Upsilon _{\theta \chi _{0},\varpi _{p}}=-\sum_{l=1}^{4}H^{\left( l\right)
}\left( -i\hat{\kappa}_{0}\right)  \\
\boldsymbol{\boldsymbol{\theta }^{T}\cdot \sigma }_{\chi _{0},\varpi
_{p}}\left( -i\hat{\kappa}_{0},\omega _{l}\right) \cdot \boldsymbol{\theta 
\boldsymbol{\theta }^{T}\cdot \tilde{\varsigma}}_{\chi _{0},\varpi
_{p}}\left( i\hat{\kappa}_{0}\right) \cdot \boldsymbol{\theta } \\
-\sum_{l,i=1}^{4}\frac{1}{\hat{\kappa}_{0}+iw_{i}^{\left( l\right) }}\text{%
\texttt{Res}}_{w=-w_{i}^{\left( l\right) }}\left[ H^{\left( l\right) }\left(
w\right) \right]  \\
\boldsymbol{\boldsymbol{\theta }^{T}\cdot \tilde{\sigma}}_{\chi _{0},\varpi
_{p}}\left( -w_{i}^{\left( l\right) },\omega _{l}\right) \cdot \boldsymbol{%
\theta \boldsymbol{\theta }^{T}\cdot \tilde{\varsigma}}_{\chi _{0},\varpi
_{p}}\left( w_{i}^{\left( l\right) }\right) \cdot \boldsymbol{\theta }
\end{multline*}%
with $\omega _{l}$ the frequencies of (\ref{sigmai}) and $w_{i}^{\left(
l\right) }$ the poles of $H^{\left( l\right) }\left( w\right) $ in the upper
complex $w$--plane,%
\begin{eqnarray*}
w_{i}^{\left( 1\right) } &=&\left\{ 2\omega _{+},~\omega _{+}-\Omega
_{+},~\omega _{+}-\Omega _{-},~\omega _{+}+\omega _{-}\right\}  \\
w_{i}^{\left( 2\right) } &=&\left\{ 2\omega _{-},~\omega _{-}-\Omega
_{-},~\omega _{-}-\Omega _{+},~\omega _{-}+\omega _{+}\right\}  \\
w_{i}^{\left( 3\right) } &=&\left\{ -2\Omega _{+},~\omega _{+}-\Omega
_{+},~\omega _{-}-\Omega _{+},~-\Omega _{+}-\Omega _{-}\right\}  \\
w_{i}^{\left( 4\right) } &=&\left\{ -2\Omega _{-},~\omega _{-}-\Omega
_{-},~\omega _{+}-\Omega _{-},~-\Omega _{-}-\Omega _{+}\right\} \;.
\end{eqnarray*}

%%%%%%%%%%%%

\end{document}